# ANCoEF: Asynchronous Neuromorphic Algorithm/Hardware Co-Exploration Framework with a Fully Asynchronous Simulator

Jian Zhang, Xiang Zhang, Jingchen Huang, Jilin Zhang, and Hong Chen*
Tsinghua University
hongchen@tsinghua.edu.cn

*Abstract*—Developing asynchronous neuromorphic hardware to meet the demands of diverse real-life edge scenarios remains significant challenges. These challenges include constraints on hardware resources and power budgets while satisfying the requirements for real-time responsiveness, reliable inference accuracy, and so on. Besides, the existing system-level simulators for asynchronous neuromorphic hardware suffer from runtime limitations. To address these challenges, we propose an **A**synchronous **N**euromorphic algorithm/hardware **Co**-**E**xploration **F**ramework (ANCoEF) including multi-objective reinforcement learning (RL)-based hardware architecture optimization method, and a fully asynchronous simulator (TrueAsync) which achieves over 2× runtime speedups than the state-of-the-art (SOTA) simulator. Our experimental results show that, the RL-based hardware architecture optimization approach of ANCoEF outperforms the SOTA method by reducing 1.81× hardware energy-delay product (EDP) with 2.73× less search time on N-MNIST dataset, and the co-exploration framework of ANCoEF improves SNN accuracy by 9.72% and reduces hardware EDP by 28.85× compared to the SOTA work on DVS128Gesture dataset. Furthermore, ANCoEF framework is evaluated on external neuromorphic dataset CIFAR10-DVS, and static datasets including CIFAR10, CIFAR100, SVHN, and Tiny-ImageNet. For instance, after 26.23 *ThreadHour* of co-exploration process, the result on CIFAR10-DVS dataset achieves an SNN accuracy of 98.48% while consuming hardware EDP of 0.54 s·nJ per sample.

## I. INTRODUCTION

A growing number of neuromorphic hardware for edge artificial intelligence (AI) applications has been designed recently, such as ReckOn [1], ANP-I [2], ANP-G [3], and others in [4-7]. As we know, asynchronous circuits are event-driven, which naturally match the sparsity in both space and time of spikes in spiking neural network (SNN) models. Therefore, most neuromorphic hardware for edge AI applications is designed with asynchronous circuits to effectively save energy [8]. Asynchronous neuromorphic hardware has made great progress in various energy-efficient real-time edge AI applications such as image recognition.

As it is widely recognized, edge AI deployments face great challenges in meeting energy-efficient, area-constrained, high-accuracy, and low-latency requirements for specific applications. As a result, application specific integrated circuit (ASIC) designs encounter difficulties in efficiently optimizing architectures within the expansive design space, including a wide range of numerical and non-numerical design options, such as the number of processing elements (PEs), buffer size, topology strategies, connectivity between neurons, mapping strategies and so on.

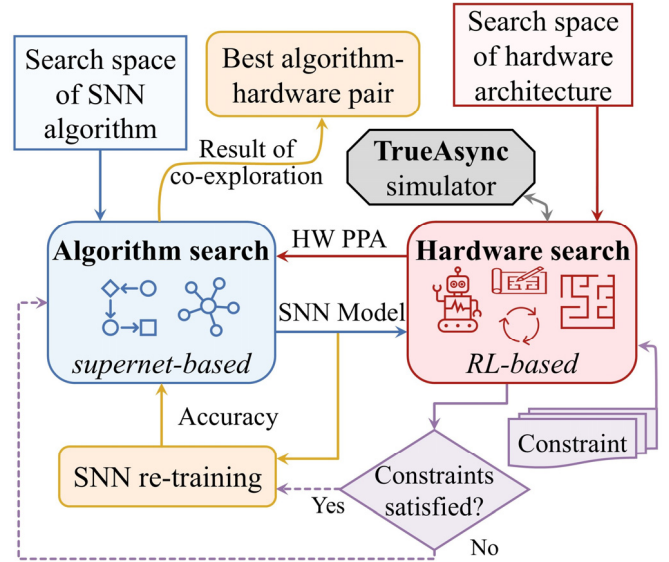

Fig. 1. Workflow of ANCoEF with supernet-based algorithm search method, RL-based hardware search method, and TrueAsync simulator.

In order to explore the large hardware design space and improve performance for deep learning neural networks (DNNs), previous researchers have successfully adopted machine learning for neural network architecture search (NAS) and hardware architecture search (HW-NAS), such as HW-NAS methods in [9-11] and algorithm/hardware co-exploration frameworks in [12-14]. However, these approaches are not suitable for optimizing SNN applications due to different hardware structures and date representations [8]. Recently, Jian Zhang et al. propose ANAS, the first HW-NAS method for asynchronous neuromorphic hardware, which uses an evolutionary algorithm to optimize both numerical and non-numerical hardware design space [8]. Nevertheless, the evolutionary algorithm used in ANAS limits the search efficiency, and the ANAS method only searches the hardware architecture design space without joint optimization of both algorithm and hardware architectures.

Besides, to evaluate the performance of asynchronous neuromorphic hardware architectures produced by these search methods, Jian Zhang et al. propose CanMore, a configurable asynchronous neuromorphic simulator, which is developed based on the cycle-accurate GEM5 architecture [8]. The CanMore simulator can only divide a synchronous cycle into several ticks and transition the states of simulated circuits tick by tick, where a smaller tick duration implies a finer resolution of events in the system. However, this cycle-accurate method for simulating asynchronous circuits requires long simulation runtime and leads to low searching efficiency consequently. To address these issues, we put forward an asynchronous neuromorphic algorithm/hardware architecture co-exploration framework (ANCoEF), and a fully asynchronous neuromorphic hardware simulator (TrueAsync). Our contributions are as follows:

• We propose a novel asynchronous neuromorphic algorithm/hardware architecture co-exploration framework (ANCoEF), which searches SNN algorithms and their hardware architectures jointly. With ANCoEF, we are able to optimize SNN accuracy and hardware performance, power, and area utilization (PPA) at the same time to meet the requirements of diverse edge AI applications.

• We put forward TrueAsync, a novel system-level simulator for asynchronous neuromorphic hardware based on the actor models. With these models in Akka.NET [15] libraries, we construct a concurrent and fully asynchronous system with handshake protocols to match asynchronous circuits in TrueAsync simulator. In comparison with the state-of-the-art (SOTA) CanMore simulator, TrueAsync achieves more than 2× simulation runtime speedups due to high concurrency, which improves the searching efficiency significantly.

• We utilize a reinforcement learning (RL)-based multi-objective optimization approach to explore the design space of asynchronous neuromorphic hardware, enhancing both effectiveness and efficiency. Our experiments show that the RL-based hardware architecture search approach of ANCoEF outperforms the evolutionary algorithm-based method of ANAS [8] by reducing 1.81× hardware energy-delay product (EDP) with 2.73× less search time on N-MNIST [25] dataset.

---

\* Hong Chen is the corresponding author.

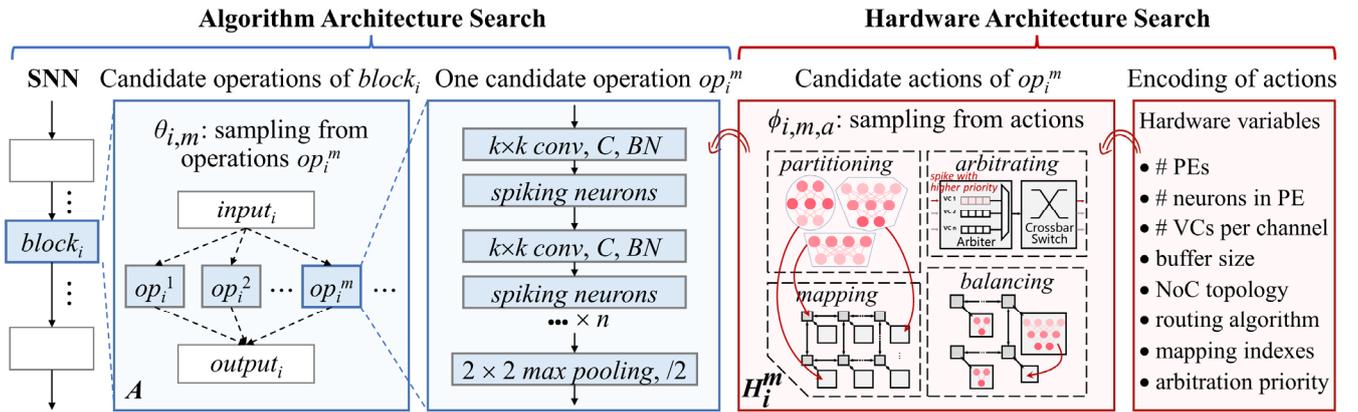

Fig. 2. ANCoEF search space that includes algorithm and hardware architecture design options.

We verify the effectiveness of ANCoEF through comprehensive experiments on various datasets, including static datasets CIFAR10 [21], CIFAR100 [21], SVHN [23], and Tiny-ImageNet [24], as well as neuromorphic datasets DVS128Gesture [20] and CIFAR10-DVS [22]. The results indicate that ANCoEF improves SNN accuracy by 9.72% and reduces hardware EDP by 28.85× on DVS128Gesture dataset compared with the SOTA search method in [8]. Additional datasets are also evaluated for various optimization objectives to establish our benchmarks.

## II. ANCoEF Co-Exploration Framework

The exploration objective of ANCoEF is to find an optimal SNN algorithm and hardware architecture pair, including SNN algorithm $\alpha$ in the search space $A$ and hardware architecture $\eta$ in the search space $H$, so that the SNN accuracy is maximized while meeting the hardware PPA target including latency, energy consumption, and area utilization:

$$\max_{A,H} Accu(\alpha, \eta) \quad (1)$$

$$\text{s.t. } L(\alpha, \eta) \leq T_{\text{latency}}, E(\alpha, \eta) \leq T_{\text{energy}}, A(\alpha, \eta) \leq T_{\text{area}} \quad (2)$$

where $Accu(\alpha, \eta)$ stands for the SNN accuracy with the architecture $\alpha$ and $\eta$, and $L$, $E$, $A$, $T$ represent the hardware latency, energy consumption, area utilization, and target constraints respectively.

Figure 1 illustrates the ANCoEF and its optimization flow. We use two algorithms for co-exploration, the supernet-based algorithm search method [16] and the RL-based hardware search method. The algorithm search method first trains a super-network to search the complex design space of SNNs, subsequently generating several candidate SNNs. These SNN candidates are then analyzed by the hardware search method, which concentrates on optimizing the hardware architectures concerning hardware performance such as PPA. As the full training of the network is more computationally expensive than hardware optimization, ANCoEF avoids unnecessary training of the candidate SNNs by partially training them first, then optimizing them with hardware search method to meet the hardware PPA target requirement, and abandoning the candidate SNNs which do not meet the PPA target, thus achieving high co-exploration efficiency. When the hardware search method finds the optimal architecture that satisfies all constraints, the SNN will be fully trained to further improve its accuracy. The architecture with the highest accuracy while meeting the hardware PPA target is regarded as the best SNN algorithm and hardware architecture pair of the co-exploration result.

### A. ANCoEF Search Space

Figure 2 shows our comprehensive search space: the blue blocks represent the SNN algorithm search space $A$, while the red blocks stand for the hardware architecture search space $H$. Each SNN is constructed with $N$ fundamental building blocks, denoted as $block_i$, where $1 \leq i \leq N$. Within the $i$-th block, there exist $M$ candidate operations, denoted as $op_i^m$ with $1 \leq m \leq M$. These operations are constructed with convolutional layers, spiking neuron layers, and pooling layers. To enhance the

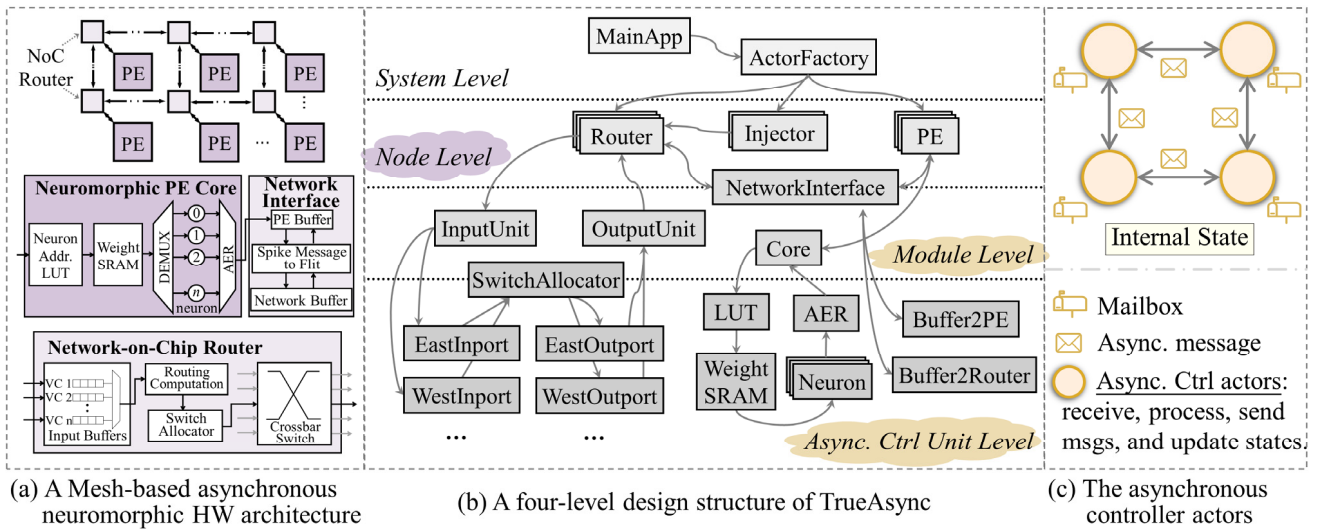

(a) A Mesh-based asynchronous neuromorphic HW architecture
(b) A four-level design structure of TrueAsync
(c) The asynchronous controller actors

Fig. 3. Architecture of the TrueAsync simulator.

searching efficiency, we delete the hardware-unfriendly design choices, such as the average pooling layer and parametric leaky integrate-and-fire (PLIF) spiking neuron model.

Similarly, in the hardware architecture search space, we abandon some hardware design choices that lead to resource wastage, such as the number of neurons in process elements (PEs) that does not align with $2^n$ ($n$=1,2,3…). This is due to the unused spiking address bits in lookup table (LUT) units, weight SRAM, address-event representation (AER) units, and network-on-chip (NoC) flit, leading to the waste of hardware resources. Moreover, we encode the hardware search space into a series of actions including partitioning, mapping, balancing, arbitrating, and altering. Hence, transitioning between two hardware architectures becomes feasible through a list of these actions, creating a diverse hardware search space $H_i^m$ for each $op_i^m$.

### B. Multi-Objective RL-Based Optimization Method

The search space of asynchronous neuromorphic hardware architectures poses great challenges because of the non-numerical design space, such as connectivity between neurons and mapping strategies. To solve this problem, we select the optimal operations in the decision process including partitioning, mapping, balancing, arbitrating, and altering, which match the actions in the RL algorithm. Besides, Q-learning reinforcement algorithm is adopted in the optimization process for hardware architecture search, since RL algorithms are widely used in creating the best architecture efficiently in recent years [14].

In decision processes, RL algorithms outperform evolutionary algorithms as RL agent takes actions following its policy instead of optimizing parameters directly. Besides, RL algorithms are not necessarily optimized from the beginning for every new application like evolutionary algorithms because RL algorithms learn from the diverse design spaces across different applications. As a result, RL algorithms outperform evolutionary algorithms in both searching effectiveness and efficiency in searching asynchronous neuromorphic hardware architecture.

In the multi-objective RL-based optimization method, we put forward a reward function $R$ to maximize SNN accuracy while meeting the hardware PPA target, including the latency of timesteps $L$, energy consumption $E$, and area utilization $A$. These performance metrics are collectively used in the reward function $R$:

$$R = Accu(\alpha, \eta) \times [\frac{L(\alpha,\eta)}{T_{latency}}]^{\omega_0} \times [\frac{E(\alpha,\eta)}{T_{energy}}]^{\omega_1} \times [\frac{A(\alpha,\eta)}{T_{area}}]^{\omega_2} \quad (3)$$

where $\omega_0$, $\omega_1$, and $\omega_2$ are the weight factors:

$$\omega_i = \begin{cases} p_i, & \text{if } PPA(\alpha, \eta) \text{ satisfies } Target \\ q_i, & \text{otherwise} \end{cases} \quad (4)$$

where hyperparameters $p_i$ and $q_i$ are application-specific constants used to define the optimization objective. Note that smaller values of $L$, $E$, $A$, and $Target$ indicate better hardware PPA. For example, when the hyperparameters $p_0 = 0$, $q_0 = -1$, we use accuracy as the objective if the simulated latency satisfies the target $T_{latency}$, otherwise, we penalize the objective value by generating a low reward $R$. When $p_0 = q_0 = -0.07$, we optimize SNN accuracy and latency $L$ jointly. Moreover, we put more attention on latency $L$ in optimization process when setting $p_0 = q_0 = -0.02$ and smaller $T_{latency}$. Both the optimization of energy consumption $E$ and area utilization $A$ follow the same rules.

Besides obtaining valid reward value from the environment during optimization, another essential task for RL agents is to select the appropriate actions that maximize rewards. We encode environment variables into agent states that determine the actions with the detail analysis tool in the TrueAsync simulator. Our encoding process factors in multiple dimensions such as AER congestion, NoC traffic congestion, total routing hops, and so on. As a result, the RL agent is able to effectively choose actions to maximize rewards in few optimization iterations.

### III. TRUEASYNC SYSTEM-LEVEL SIMULATOR

In this section, we introduce the TrueAsync system-level simulator, which is able to model asynchronous neuromorphic hardware systems using the actor model from Akka.NET libraries.

### A. A Fully Asynchronous System Using Actor Models

As we know, asynchronous circuits have two-phase and four-phase handshake protocols, and different styles such as bundled-data (BD) and quasi-delay insensitive (QDI) circuits, but they all share the characteristic that a design is partitioned into a collection of concurrently operating hardware modules that communicate with each other using well-defined protocols on wire bundles channels [17]. In designing TrueAsync simulator, we use the actor model in Akka.NET [15], a multithreaded asynchronous programming model, to represent asynchronous circuits.

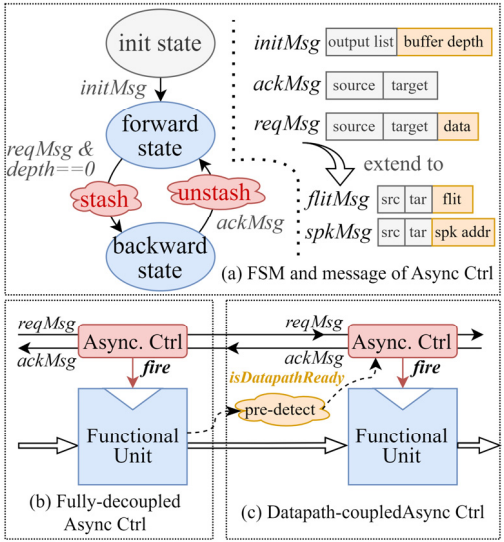

TABLE I. NoC Router Parameters in TSMC 180nm Technology

| Category | Latency | | Leakage Power | Area Utilization |
|---|---|---|---|---|
| | Forward | Backward | | |
| Input Unit | 1.2 ns | 1.5 ns | 0.063 mW | 20547 um$^2$ |
| Output Unit | 1.6 ns | 2.0 ns | 0.044 mW | 14536 um$^2$ |
| Switch Allocator | 1.9 ns | 2.4 ns | 0.031 mW | 10764 um$^2$ |

TABLE II. TrueAsync Runtime Compared with CanMore [8]

| Application | Runtime (*ThreadHour*[1]) | | Speedup |
|---|---|---|---|
| | CanMore | TrueAsync | |
| MLP-MNIST[2] | 0.98 | 0.48 | 2.01× |
| CSNN-CIFAR10[3] | 66.67 | 4.22 | 15.8× |

[1] *ThreadHour* is the product of runtime in hours and the number of threads on the evaluation platform. Two simulators are evaluated on the same platform with 40 threads.
[2] We use the same fully connected network structure FC (784, 512, 10) and 100 timesteps.
[3] We use the same convolutional network structure reported in [8] and 4 timesteps.

Fig. 4. Actor model-based asynchronous controllers.

Figure 3 (a) illustrates a mesh-based asynchronous neuromorphic hardware, consisting of numerous PE cores, NoC routers, and network interfaces bridging corresponding routers and PEs. To simulate these architectures, we employ a four-level abstraction in TrueAsync, including system level, node level, module level, and asynchronous controller (Async Ctrl) unit level, to create a comprehensive and structured simulation environment, as shown in Figure 3 (b). Async Ctrl unit level is designed to model and simulate dataflow behaviors of asynchronous neuromorphic hardware. Hundreds of actors are created at this level to implement circuit functions, each of which is based on an Async Ctrl actor to model asynchronous circuit dataflow.

*B. Asynchronous Controller Actor*

In order to model and simulate the dataflow behaviors of asynchronous circuits, we design Async Ctrl actors, which are developed based on finite-state machines (FSM) and separated from the data path in the asynchronous pipelines.

**FSM-based Async Ctrl**: As shown in Figure 4 (a), Async Ctrl actors have three different working states: *init state*, *forward state*, and *backward state*. The transition from the *init state* to the *forward state* occurs when the Async Ctrl receives *initMsg* with output list and buffer depth. In the *forward state*, the Async Ctrl is responsible for asynchronous communication using *reqMsg* and *ackMsg*. The transition from the *forward state* to the *backward state* happens when the buffer depth of the Async Ctrl is zero. In the *backward state*, the Async Ctrl stashes incoming request messages (*reqMsg*) and transitions to the *forward state* when receiving an acknowledgment message (*ackMsg*). With the FSM-based feature of the Async Ctrl actors, our TrueAsync simulator is able to model various behaviors and delays of asynchronous circuits in the forward and backward states.

**Fully datapath-decoupled Async Ctrl**: As shown in Figure 4 (b), Async Ctrl actors model the asynchronous circuit dataflow by separating the handshake control and data paths. In most cases, the handshake control and data paths are fully decoupled, and the controller generates the *fire* signal as the local clock to trigger the registers in this pipeline stage. To implement circuit functions using Async Ctrl, functional units are required to extend the standard Async Ctrl actor and override the main function which is responsible for processing data within the *reqMsg*. For example, to implement the east input port (Figure 4 (b)) using the Async Ctrl actor, a functional unit that extends the Async Ctrl actor is required, and it must override the main function, where the data within the asynchronous request message are the input of FIFO, and the output of FIFO are required to return, modeling the east input port function of the circuit. Moreover, the extended functional units are unable to directly receive and send asynchronous messages while waiting for the *fire* signal from the handshake controller, modeling the structure of asynchronous circuit pipelines.

**Datapath-coupled Async Ctrl**: In situations, such as in feedback or cycling circuits, the data path is required to be involved in the handshake control path to jointly generate the *fire* signal. Therefore, we design data path-coupled Async Ctrl actors, in which the main function of data path returns an *isDatapathReady* signal to the handshake control path. When a data path-coupled Async Ctrl actor receives the *isDatapathReady* signal, it starts to send the acknowledgment message (*ackMsg*) to the upstream pipeline stage and send the request message (*reqMsg*) to the downstream pipeline stage.

*C. Obtaining Parameters of Technology for TrueAsync from ASIC Design*

To make use of the real parameters of a specific technology for our TrueAsync simulator, we design a register transfer level (RTL) asynchronous NoC router with TSMC 180nm technology. The router has 5 ports, 4 virtual channels, and 8 stages FIFO for each port, using Click [18]-based asynchronous pipelines. From the synthesis results of the asynchronous NoC router by commercial EDA tools, we are able to obtain the real hardware PPA parameters of the router, as shown in Table I. These parameters are then injected into the NoC router modeling simulated by TrueAsync. We record the switching activity of modules in TrueAsync and add the leakage power from the synthesis results to estimate the total energy consumption.

TABLE III. RESULTS OF HARDWARE ARCHITECTURE SEARCH COMPARED WITH ANAS[8]

| Model [8] | Topology | ANAS[8] EDP (s·nJ) | ANAS[8] Search ThreadHour | ANCoEF EDP (s·nJ) | ANCoEF EDP Reduction | ANCoEF Search ThreadHour | ANCoEF Time Saving |
|---|---|---|---|---|---|---|---|
| S-256 | FC (128, 64, 64) | 9e7 | 20 | 4.98e7 | **1.81×** | 7.33 | **2.73×** |
| S-512 | FC (256, 128, 128) | 7e6 | 104 | 1.75e7 | -- | 17.5 | **5.94×** |
| S-1024 | FC (512, 256, 256) | 2e8 | 284.67 | 1.42e8 | **1.41×** | 61 | **4.67×** |
| S-2048 | FC (1024, 512, 512) | 2e8 | 6666.67 | 2e8 | -- | 80.17 | **83.16×** |

TABLE IV. RESULTS OF ALGORITHM/HARDWARE ARCHITECTURE CO-EXPLORATION COMPARED WITH ANAS [8] AND OUR BENCHMARKS

| Dataset | Model[1] | Timestep | Accuracy | Energy[2] (μJ) | Latency[2] (μs) | Area (mm[2]) | EDP[2] (s·nJ) | Search Time (ThreadHour) |
|---|---|---|---|---|---|---|---|---|
| MNIST[25] | ANAS-MNet | 100 | 98.81% | -- | -- | -- | 1 | 22.67 |
|  | ANCoEF-MNet | 4 | 98.42% | 9.05 | 29.99 | 0.32 | **0.27** | **7.96** |
| DVS128Gesture [20] | ANAS-DGNet | 16 | 85.42% | -- | -- | -- | 30 | 6666.7 |
|  | ANCoEF-DGNet-A | 5 | 92.71% | **6.47** | **28.46** | 3.22 | **0.184** | **31.52** |
|  | ANCoEF-DGNet-B | 16 | **95.14%** | 14.93 | 69.56 | 7.35 | 1.04 | 56.43 |
| CIFAR10[21] | ANAS-CIFAR10Net | 16 | 76.74% | -- | -- | -- | 4 | 20000 |
|  | **ANCoEF-Net-16** | 16 | 86.11% | **61.39** | **138.09** | 4.73 | **8.48** | **63.28** |
|  | **ANCoEF-Net-128** | 16 | **90.12%** | 653.93 | 1140.8 | 19.05 | 745.97 | 98.32 |
| CIFAR100[21] | **ANCoEF-Net-16** | 16 | 54.35% | **209.06** | **520.66** | 4.73 | **108.85** | **69.65** |
|  | **ANCoEF-Net-128** | 16 | **61.09%** | 1286.3 | 3004.6 | 19.05 | 3864.9 | 116.89 |
| CIFAR10-DVS[22] | **ANCoEF-Net-16** | 4 | 91.35% | **11.13** | **38.93** | **4.73** | **0.43** | **14.73** |
|  |  | 16 | **98.48%** | 12.27 | 44.38 | **4.73** | 0.54 | 26.23 |
| SVHN[23] | **ANCoEF-Net-64** | 16 | 95.32% | 318.14 | 549.8 | 11.3 | 174.91 | 48.24 |
| Tiny-ImageNet[24] | **ANCoEF-Net-64** | 16 | 46.79% | 350.21 | 785.08 | 11.3 | 274.94 | 147.64 |

[1] ANAS-MNet: FC(512). ANCoEF-MNet: FC(256,128). ANAS-DGNet: ConvStem-FC(800). ANCoEF-DGNet-A: ConvStem-4×{C48K3-M2}-FC(512). ANCoEF-DGNet-B: ConvStem-4×{C16K3-M2}-FC(512). ANAS-CIFAR10Net: C16K5-M2-C32K5-M2-FC(800,128). ANCoEF-Net-n: ConvStem-{C{n}K5}×2-M2-{C{2n}K5}×2-M2-{C{4n}K3}×2-M2-{C{4n}K5}×2-M2-FC(1024), where n stands for the number of initial input channels.

[2] Average results of one frame on each dataset.

## IV. EVALUATION

In this section, we compare the runtime of TrueAsync with the SOTA simulator: CanMore, evaluate the reinforcement learning-based hardware search method of ANCoEF by comparing it with the evolutionary algorithm-based search method in ANAS [8], and demonstrate algorithm/hardware co-exploration results of ANCoEF on various static datasets and neuromorphic datasets. All experiments are performed on two Intel(R) Xeon(R) Silver 4210R CPUs @ 2.40 GHz containing 20 cores and 40 threads in total, and four GeForce RTX 3090 GPUs.

### A. Simulation Runtime Comparison with the SOTA Work

To verify TrueAsync simulator, we compare its average runtime with that of the CanMore simulator, using identical network structures and algorithm timesteps on the same hardware platform. The runtime of the simulators is denoted as *ThreadHour*, which is the product of runtime in hours and the number of threads on the evaluation platform. The comparison results are illustrated in Table II, from which we find that TrueAsync achieves a 2.01× runtime reduction for the MLP-MNIST application and a 15.8× runtime reduction for the CSNN-CIFAR10 application. These reductions in simulation time are due to the concurrent and fully asynchronous system with handshake protocols in our TrueAsync simulator, leveraging multicore performance with high concurrency. In contrast, the CanMore simulator is cycle-accurate without concurrent and parallel computation abilities.

### B. Evaluation of ANCoEF RL-based HW-NAS Method

To verify our RL-based hardware architecture search method, we compare the EDP of the optimal hardware architectures with that of ANAS [8]. Applications of various scales are pretrained on N-MNIST [25] dataset to perform hardware architecture search. Switching activity interchanging format (SAIF) is used to estimate the energy consumption for both CanMore and TrueAsync based on the detailed analysis of Neurogrid chip reported in [19]. The comparison results are shown in Table III and Figure 5.

Compared to the evolutionary algorithm-based search of ANAS, our RL-based hardware architecture search method of ANCoEF achieves 1.81× and 1.41× hardware EDP reductions for S-256 and S-1024 applications respectively, as illustrated in Figure 5. More importantly, as shown in Table III, our method requires less search time to obtain the optimal hardware architecture, saving from 2.73× to 83.16× search time compared to ANAS method, enabling efficient optimization of large-scale applications on complex datasets such as CIFAR100 [21] and Tiny-ImageNet [24].

### C. Evaluation of ANCoEF Co-Exploration Framework

Combining the algorithm and hardware architecture search processes, ANCoEF is able to find the optimal SNN algorithm-hardware architecture pairs on various static and neuromorphic datasets, enhancing hardware while maintaining SNN accuracy. Table IV shows the comparison results between our method and ANAS [8] method on various datasets, including MNIST [25], DVS128Gesture [20],

CIFAR10 [21], CIFAR100 [21], CIFAR10-DVS [22], SVHN [23] and Tiny-ImageNet [24] datasets.

On MNIST dataset, the SNN evaluated by ANAS method adopts 100 timesteps, while the optimal architecture searched by ANCoEF has 4 timesteps and fewer neurons, which results in a 3.7× reduction of hardware EDP per frame at a cost of only 0.39% drop in SNN accuracy. On DVS128Gesture dataset, the simple fully connected network used in ANAS leads to a low SNN accuracy of 85.42%. After co-exploration by ANCoEF, ANCoEF-DGNet-A with 4 convolutional layers and 2 fully connected layers achieves an accuracy of 92.71% and consumes 0.184 s·nJ hardware EDP per frame, which is improved by 163× compared to that of ANAS-DGNet. ANCoEF-DGNet-B further improves by 2.43% in SNN accuracy with fewer timesteps. On CIFAR10 dataset, the 4-layer network used in ANAS has an accuracy of 76.74%, while the 9-layer network ANCoEF-Net-128 searched by ANCoEF achieves 90.12% accuracy, improving by 13.38%. Meanwhile, ANCoEF-Net-16 further reduces hardware EDP by 87.97× with 86.11% SNN accuracy by cutting down the channel size. This searching process takes 63.28 *ThreadHours*, only 0.3% of the search time required by ANAS on CIFAR10 dataset.

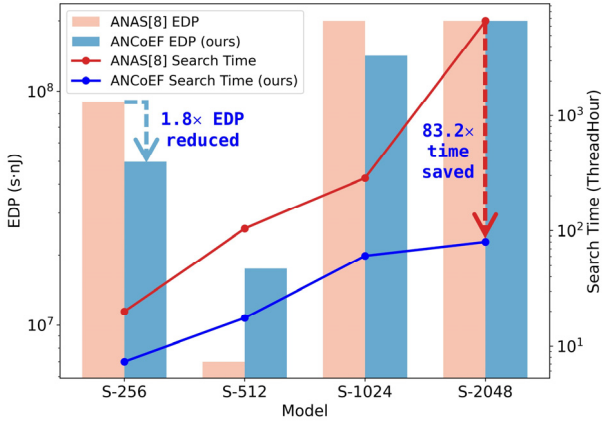

Fig. 5. Results of hardware architecture search compared with ANAS[8].

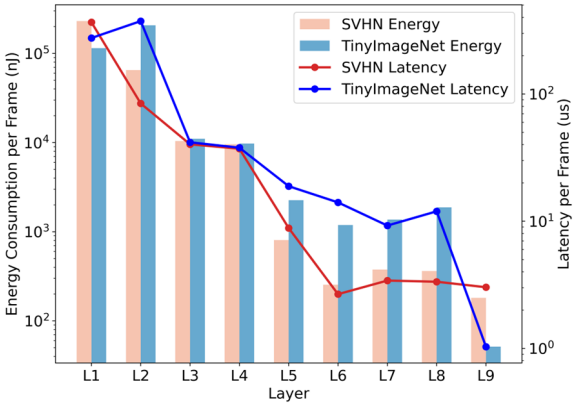

Fig. 6. Layer-wise co-exploration results by ANCoEF on SVHN and Tiny-ImageNet datasets.

Furthermore, we evaluate the multi-objective optimization method on more complex datasets, including CIFAR100, SVHN, and Tiny-ImageNet static datasets, and CIFAR10-DVS neuromorphic dataset. On CIFAR100 and CIFAR10-DVS datasets, we set $p_i = q_i = -0.07$ in the equation (3) and (4) to make the reward jointly optimize the hardware PPA and SNN accuracy, and set $p_i = q_i = -0.02$ to search for the architectures with lower hardware PPA. For instance, on CIFAR100 dataset, ANCoEF-Net-16 reduces hardware EDP by 35.51× per frame with fewer channels compared with ANCoEF-Net-128. On CIFAR10-DVS dataset, the model with the same network structure but fewer timesteps reduces hardware EDP by 1.26× per frame.

Figure 6 depicts how the hardware EDP per frame of convolutional and fully connected layers of SNNs varies on SVHN and Tiny-ImageNet datasets using ANCoEF-Net-64 network model. We observe that the first and second convolutional layers have the highest energy consumption and latency in comparison to the other layers. This is because more neurons are implemented in these two layers with larger input feature dimensions. Despite utilizing the same ANCoEF-Net-64 network model, the two SNNs on different datasets result in different hardware PPA because of differences in the number of spikes between layers. Therefore, the model on the complex Tiny-ImageNet dataset generates more spikes resulting in increased hardware energy consumption and latency.

## V. CONCLUSION

In this paper, we introduce the ANCoEF co-exploration framework for optimizing SNN algorithm and asynchronous neuromorphic hardware architectures simultaneously. This framework leverages a multi-objective reinforcement learning algorithm to optimize hardware architectures. To enhance searching efficiency, we develop TrueAsync, a fully asynchronous simulator based on Akka actor models, capable of modeling and simulating asynchronous neuromorphic hardware architectures. TrueAsync provides more than 2× runtime speedups compared to the SOTA simulator CanMore in [8]. With TrueAsync, ANCoEF rivals the SOTA HW-NAS method of ANAS[8], achieving a 1.81× reduction in hardware EDP and over a 2.73× reduction in search time for hardware architecture search only. Moreover, through the co-exploration process of SNN algorithms and hardware architectures by ANCoEF, hardware EDP is further reduced by 3.7× on MNIST dataset and 163× on DVS128Gesture dataset. On CIFAR10 dataset, we use only 0.3% of the search time to find the optimal architecture, achieving 8.48 s·nJ hardware EDP per frame and a 9.37% increase of SNN accuracy compared with the searched architecture of ANAS. Finally, we verify the multi-objective co-exploration optimization of ANCoEF and establish our benchmarks on additional datasets such as CIFAR100, SVHN, Tiny-ImageNet static datasets, and CIFAR10-DVS neuromorphic dataset.